\documentclass[twocolumn,prb,showpacs]{revtex4}
\usepackage{graphicx}
\usepackage{dcolumn}
\usepackage{bm}
\usepackage{amsmath}

\begin{document}

\preprint{}
\title{Charge pumping due to triplet vector chirality in ferromagnet/triplet superconductor junctions}
\author{Takehito Yokoyama}
\affiliation{Department of Physics, Tokyo Institute of Technology, Tokyo 152-8551,
Japan 
%\\ $^2$Department of Physics, Tokyo Metropolitan University, Hachioji, Tokyo 192-0397, Japan 
}
\date{\today}

\begin{abstract}
We investigate charge pumping in ferromagnet/triplet superconductor junctions where the magnetization of the ferromagnet is inhomogeneous and dynamical. It is shown that charge current is pumped due to the coupling of the localized spins with triplet vector chirality, vector chirality formed by the triplet vector of Cooper pairing. Physical mechanism of the charge pumping is also discussed. 
%This mechanism does not rely on spin-orbit coupling in contrast to the previous mechanism of charge pumping by magnetization dynamics. 
\end{abstract}

\maketitle

%\affiliation{$^1$Department of Physics, Tokyo Institute of Technology, 2-12-1 Ookayama, Meguro-ku, Tokyo 152-8551, Japan \\

Spintronics aims to manipulate electron's spin electrically or control electron transport magnetically. Spin current, a flow of spin angular momentum carried by electrons, is one of central concepts in spintronics. There are several methods to create spin current. Among them, spin current generation by precession of magnetization, called \textit{spin pumping}, has been widely used.\cite{Silsbee,Tserkovnyak,Tserkovnyak2} 
%To detect spin current, one may convert it into electric signal by the inverse spin Hall effect \cite{Saitoh,Valenzuela} where the conversion occurs due to spin-orbit coupling. 
Pumped spin current can be converted into electric signal by the inverse spin Hall effect \cite{Saitoh,Valenzuela} where the conversion occurs due to spin-orbit coupling. In this way, charge current is generated by magnetization dynamics. 
In this paper, we address the problem of conversion of magnetization dynamics into charge current in ferromagnet/triplet superconductor junctions.

Spintronics is also relevant to triplet superconductor since Cooper pairs have spin degree of freedom.\cite{Sigrist} 
To date, several materials such as Sr$_2$RuO$_4$\cite{Maeno,Mackenzie} and UPt$_3$\cite{Brison} have been identified as triplet superconductors. 
When triplet superconductivity and magnetism coexist, one may expect remarkable interplay between them since both have spin degree of freedom. In fact, it has been shown that, in ferromagnetic Josephson junctions with triplet superconductors, the interplay between triplet superconductivity and magnetism is manifested in the Josephson current. \cite{Kastening,Brydon} This interplay has also been investigated in ferromagnetic superconductors where ferromagnetism and  triplet superconductivity coexist in the bulk state.\cite{Brataas,Gronsleth}
Also, Josephson junctions composed of triplet superconductor have been fabricated to identify the pairing symmetry of triplet superconductors.\cite{Nelson,Strand,Strand2}

In this paper, we study charge pumping in ferromagnet/triplet superconductor junctions where the magnetization of the ferromagnet is inhomogeneous and dynamical. It is shown that charge current is pumped due to the coupling of the localized spins with triplet vector chirality, vector chirality formed by the triplet vector of Cooper pairing. This can be also represented as a coupling between the pumed spin current  and the  triplet vector chirality. 
Physical mechanism of the charge pumping is also presented. 
This mechanism does not rely on spin-orbit coupling in contrast to the previous mechanism of charge pumping by magnetization dynamics\cite{Saitoh,Valenzuela}. 

%%%%%%%%%%%%%%%%%%%%% Formulation
\begin{figure}[tbp]
\begin{center}
\scalebox{0.8}{
\includegraphics[width=8.50cm,clip]{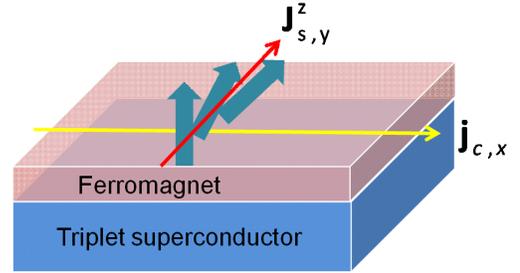}
}
\end{center}
\caption{(Color online) Schematic of the model. Pumped charge current $j_{c,x} ({\bf{x}},t)$ and the pumped spin current $j_{s,y}^z ({\bf{x}},t) \propto \nabla _y \dot{{\bf{S}}}^z ({\bf{x}},t)$ for helical $p$-wave superconductivity  (Eq.(\ref{jcex})). }
\label{fig1}
\end{figure}
%%%%%%%%%%%%%%%%%%%%%%%%

We consider a ferromagnet/triplet superconductor junction where the magnetization of the ferromagnet is inhomogeneous and dynamical as shown in Fig. \ref{fig1}. The Hamiltonian of the superconductor and the ferromagnet are given by  $H_S  = H_0  + H_\Delta $ and $H_F  = H_0  + H_{ex}$, respectively. 
The $H_0$, $H_\Delta$ and  $H_{ex}$ represent the kinetic energy, the superconducting order, and the exchange interaction between the conducting electron and the localized spins, respectively:
\begin{eqnarray}
 H_0  = \sum\limits_{\bf{k}} {\phi _{\bf{k}}^\dag  \xi {\sigma _0  \otimes } \tau _3 \phi _{\bf{k}}^{} }  ,\\ H_\Delta   = \sum\limits_{\bf{k}} {\phi _{\bf{k}}^\dag  \left[ {({\bf{d}}_1  \cdot {\bm{\sigma }}) \otimes \tau _1  + ({\bf{d}}_2  \cdot {\bm{\sigma }}) \otimes \tau _2 } \right]\phi _{\bf{k}}^{} },
  \\ 
H_{ex}  =  - J\sum\limits_{{\bf{k}},{\bf{q}}} {(\phi _{{\bf{k}} - {\bf{q}}}^\dag  {\bm{\sigma }} \otimes \tau _0 \phi _{\bf{k}}^{} ) \cdot {\bf{S}}_{\bf{q}}^{} } (t)
 \label{hex}
\end{eqnarray}
with $\phi _{\bf{k}}^\dag   = (c_{{\bf{k}} \uparrow }^\dag  ,c_{{\bf{k}} \downarrow }^\dag  ,ic_{ - {\bf{k}} \downarrow }^{} , - ic_{ - {\bf{k}} \uparrow }^{} )$ and $\xi  = \frac{{\hbar ^2 k^2 }}{{2m}} - \varepsilon _F$. Here, $\sigma$ and $\tau$ are Pauli matrices in spin and Nambu spaces, respectively. Also, $\varepsilon _F$, ${\bf{d}}_j (j=1,2)$, $J$, and $\bf{S}$ are the Fermi energy, the $\bf{d}$-vector of triplet pairing, the exchange coupling, and the localized spins, respectively. 
The localized spins depend on space and time, but we consider only slowly varying case. The dynamics of spins can be driven by applying magnetic field. 
We take into account $H_{ex}$ as a first order perturbation. 
Note that we adopt the basis in Ref.\cite{Ivanov} such that singlet pairing is proportional to the unit matrix in spin space. 
With the above Hamiltonians,
% the velocity operator reads 
%\begin{eqnarray}
%v_i  = \frac{\hbar }{m}k_i \tau _3  + \delta _{ij} \frac{e}{m}\nabla _j \varph\end{eqnarray}
the charge ($j_{c,i}^{} $) and spin current ($j_{s,i}^\alpha$) operators in $i$-direction read
\begin{eqnarray}
j_{c,i}^{}  =  - \frac{{e\hbar }}{m}k_i \sigma _0  \otimes \tau _0, \quad  j_{s,i}^\alpha   = \frac{{\hbar ^2 }}{{2m}}k_i  \sigma _\alpha  \otimes \tau _3
\end{eqnarray}
where $\alpha$ is the polarization direction of spin.

%\begin{figure}[tbp]
%\begin{center}
%\scalebox{0.8}{
%\includegraphics[width=10.5cm,clip]{fig3.eps}}
%\end{center}
%\caption{(Color) Normalized thermal Hall conductivity $\kappa _{xy}/\Delta$ in unit of $\frac{{k_B^2}}{h}$. Thin black lines in the upper panel represents the chemical potential as a function of $T$ for fixed electron densities. Lower panel: temperature dependence at $\mu/\Delta=0.5$. The straight line in the lower panel represents the low-temperature asymptotic behavior in  Eq.(\ref{kappa3}).}
%\label{fig3}
%\end{figure}

\begin{figure}[tbp]
\begin{center}
\scalebox{0.8}{
\includegraphics[width=6.0cm,clip]{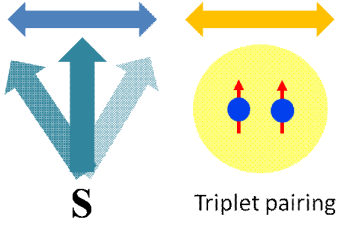}
}
\end{center}
\caption{(Color online) Mechanism of the charge pumping. The dynamics of the localized spins is transferred to the triplet Cooper pairing via the exchange coupling. The charge pumping occurs by the dynamics of the Cooper pair with charge $-2e$. }
\label{fig0}
\end{figure}

First, let us explain physical mechanism of the charge pumping intuitively. 
The charge pumping occurs in the ferromagnet by the motion of proximity induced Cooper pairs as follows. In the ferromagnet, through the exchange coupling, Eq.(\ref{hex}), the dynamics of the localized spins is transferred to the electron's spin. Since Cooper pair is formed by two electrons,  the triplet vector also acquires dynamics (triplet vector is defined in Eq.(\ref{g})). The dynamics of the triplet vector is converted into motion of the Cooper pair via the momentum dependence of the triplet vector. Then, charge pumping occurs by the vibration of the Cooper pair with charge $-2e$ as depicted in Fig. \ref{fig0}. This picture is supported by the analytical expression of the pumped charge current (Eq.(\ref{jc2})).

Now, let us calculate the charge current and give the analytical expression. We consider the unperturbed advanced Green's functions in the ferromagnet of the form 
\begin{eqnarray}
g_{{\bf{k}},\omega }^a  = g_{0,{\bf{k}},\omega }^a \sigma _0  \otimes \tau _0  + g_{3,{\bf{k}},\omega }^a \sigma _0  \otimes \tau _3 \nonumber \\ + ({\bf{f}}_{1,{\bf{k}},\omega }^a  \cdot {\bm{\sigma }}) \otimes \tau _1  + ({\bf{f}}_{2,{\bf{k}},\omega }^a  \cdot {\bm{\sigma }}) \otimes \tau _2 \label{g}
\end{eqnarray}
where $g_{0,{\bf{k}},\omega }^a$ and $g_{3,{\bf{k}},\omega }^a$ are normal Green's functions while ${\bf{f}}_{1,{\bf{k}},\omega }^a$ and ${\bf{f}}_{2,{\bf{k}},\omega }^a$ are 3D vector characterizing anomalous Green's function. \cite{comment1}
The charge current can be represented as \cite{Haug}
\begin{eqnarray}
j_{c,i} ({\bf{x}},t)  = \frac{{i\hbar ^2 e}}{{mV}}\sum\limits_{{\bf{k}},{\bf{q}}} {e^{ - i{\bf{q}} \cdot {\bf{x}}} {\rm{Tr}}k_i G_{{\bf{k}} - {\bf{q}}/2,{\bf{k}} + {\bf{q}}/2}^ <  (t,t)} 
\end{eqnarray}
where $V$ is the total volume and ${\rm{Tr}}$ is taken over spin and Nambu spaces. $G_{{\bf{k}} - {\bf{q}}/2,{\bf{k}} + {\bf{q}}/2}^ <  (t,t)$ is the lesser Green's function of the total Hamiltonian. 
Performing perturbation with respect to $H_{ex}$, we expand the lesser component using the advanced Green's functions by the Langreth theorem.\cite{Haug}
Noting that $g_{{\bf{k}},\omega }^ <   = f_\omega  \left[ {g_{{\bf{k}},\omega }^a  - (g_{{\bf{k}},\omega }^a )^\dag  } \right]$ with the lesser Green's function $g_{{\bf{k}},\omega }^ <$  and the Fermi distribution function $f_\omega$, we can compute the charge current.
The first order expansion with respect to $J$ is thus given by
\begin{widetext}
\begin{eqnarray}
 j_{c,i} ({\bf{x}},t) \cong  - \frac{{\hbar ^2 e}}{{Vm}}J\sum\limits_{{\bf{k}},{\bf{q}},\omega ,\Omega } {e^{ - i{\bf{q}} \cdot {\bf{x}} + i\Omega t} {\rm{Tr}}} k_i \left[ {g_{{\bf{k}} - {\bf{q}}/2,\omega  - \Omega /2}^{} ({\bf{S}}_{{\bf{q}},\Omega }^{}  \cdot {\bm{\sigma }}) \otimes \tau _0 g_{{\bf{k}} + {\bf{q}}/2,\omega  + \Omega /2}^{} } \right]_{}^ <  \nonumber \\ 
  \cong \frac{{4\hbar ^2 eJ}}{{Vm}}\nabla _l \dot{\bf{S}}^\alpha  ({\bf{x}},t)\sum\limits_{{\bf{k}},\omega } {f'_\omega  } k_i {\mathop{\rm Re}\nolimits} \left[ {\left( {\frac{\partial }{{\partial k_l }}{\bf{f}}_{j,{\bf{k}},\omega }^a  \times {\bf{f}}_{j,{\bf{k}},\omega }^r } \right)^\alpha  } \right] \label{jc}
\end{eqnarray}
\end{widetext}
where we assume that the frequency of spin dynamics $\Omega$ is sufficiently large such that the equilibrium current is negligible compared to the dynamical one, and the repeated indices $\alpha, l$ and $j$ are summed over ($\alpha, l=1,2,3, j=1,2$). 
Eq.(\ref{jc}) is the central result of this paper. 
From this equation, we find that coupling of the localized spins  with the triplet vector chirality in momentum space 
\begin{eqnarray}
{\frac{\partial }{{\partial k_l }}{\bf{f}}_{j,{\bf{k}},\omega }^a  \times {\bf{f}}_{j,{\bf{k}},\omega }^r }
\end{eqnarray} 
yields the pumped charge current. Therefore, in the normal state or in the absence of the triplet vector chirality, the pumped charge current vanishes.

Similarly, the spin current can be represented as 
\begin{eqnarray}
j_{s,i}^\alpha  ({\bf{x}},t) =  - \frac{{i\hbar ^3 }}{{2mV}} \nonumber \\
\times  \sum\limits_{{\bf{k}},{\bf{q}}} {e^{ - i{\bf{q}} \cdot {\bf{x}}} {\rm{Tr}}k_i  \sigma_\alpha \otimes \tau _3  G_{{\bf{k}} - {\bf{q}}/2,{\bf{k}} + {\bf{q}}/2}^ <  (t,t)}.
\end{eqnarray}
The spin current by the first order expansion with respect to $J$ can be calculated as 
\begin{widetext}
\begin{eqnarray}
 j_{s,i}^\alpha  ({\bf{x}},t) \cong \frac{{i\hbar ^3 }}{{2mV}}J\sum\limits_{{\bf{k}},{\bf{q}},\omega ,\Omega } {e^{ - i{\bf{q}} \cdot {\bf{x}} + i\Omega t} {\rm{Tr}}} k_i \sigma _\alpha \otimes \tau _3  \left[ {g_{{\bf{k}} - {\bf{q}}/2,\omega  - \Omega /2}^{} ({\bf{S}}_{{\bf{q}},\Omega }^{}  \cdot {\bm{\sigma }}) \otimes \tau _0 g_{{\bf{k}} + {\bf{q}}/2,\omega  + \Omega /2}^{} } \right]_{}^ <  \nonumber \\ 
  =  \frac{{4\hbar ^3 J}}{{Vm}}\nabla _l {\bf{S}}^\beta  ({\bf{x}},t)\sum\limits_{{\bf{k}},\omega } {f_\omega  } k_i {\mathop{\rm Im}\nolimits} \left[ {\Lambda _{{\bf{k}},\omega}^{\alpha \beta, aa} } \right] \nonumber \\ 
  - \frac{{2\hbar ^3 J}}{{Vm}}\nabla _l {\dot {\bf{S}}}^\beta  ({\bf{x}},t)\sum\limits_{{\bf{k}},\omega } {f'_\omega  } k_i \left[ {\delta _{\alpha \beta } \delta _{il} {\mathop{\rm Im}\nolimits} \left\{ { - \frac{\partial }{{\partial k_l }}g_{0,{\bf{k}},\omega }^r g_{3,{\bf{k}},\omega }^a  + g_{0,{\bf{k}},\omega }^r \frac{\partial }{{\partial k_l }}g_{3,{\bf{k}},\omega }^a } \right\} + {\mathop{\rm Re}\nolimits} \left\{ {\Lambda _{{\bf{k}},\omega }^{\alpha \beta, ra} } \right\}} \right] \label{js}
\end{eqnarray}
where 
\begin{eqnarray}
 \Lambda _{{\bf{k}},\omega }^{\alpha \beta, aa} =  - \left( {\frac{\partial }{{\partial k_l }}{\bf{f}}_{1,{\bf{k}},\omega }^a } \right)^\beta  \left( {{\bf{f}}_{2,{\bf{k}},\omega }^a } \right)^\alpha   + \left( {{\bf{f}}_{1,{\bf{k}},\omega }^a } \right)^\beta  \left( {\frac{\partial }{{\partial k_l }}{\bf{f}}_{2,{\bf{k}},\omega }^a } \right)^\alpha  
+ \delta _{\alpha \beta } \left( {\frac{\partial }{{\partial k_l }}{\bf{f}}_{1,{\bf{k}},\omega }^a  \cdot {\bf{f}}_{2,{\bf{k}},\omega }^a  - {\bf{f}}_{1,{\bf{k}},\omega }^a  \cdot \frac{\partial }{{\partial k_l }}{\bf{f}}_{2,{\bf{k}},\omega }^a } \right) \nonumber \\ 
  + \left( {\frac{\partial }{{\partial k_l }}{\bf{f}}_{2,{\bf{k}},\omega }^a } \right)^\beta  \left( {{\bf{f}}_{1,{\bf{k}},\omega }^a } \right)^\alpha   - \left( {{\bf{f}}_{2,{\bf{k}},\omega }^a } \right)^\beta  \left( {\frac{\partial }{{\partial k_l }}{\bf{f}}_{1,{\bf{k}},\omega }^a } \right)^\alpha 
\end{eqnarray}
\end{widetext}
and $\Lambda _{{\bf{k}},\omega }^{\alpha \beta, ra}$ is defined by $\Lambda _{{\bf{k}},\omega }^{\alpha \beta, aa}$ with replacing ${\bf{f}}_{1,{\bf{k}},\omega }^a$  by ${\bf{f}}_{1,{\bf{k}},\omega }^r$.
The first term in Eq.(\ref{js}) corresponds to the equilibrium spin current while the second term  describes dynamical component of the spin current, namely pumped spin current. $\Lambda _{{\bf{k}},\omega }^{\alpha \beta, aa (ra)}$ becomes zero when one of ${\bf{f}}_{1,{\bf{k}},\omega }^{a(r)}$ and ${\bf{f}}_{2,{\bf{k}},\omega }^a$ is zero vector. Then, the equilibrium spin current vanishes while the dynamical spin current can be present. 
We also see from Eq.(\ref{js}) that normal component of the pumped spin current (the term containing normal Green's functions $g_{0,{\bf{k}},\omega }^r$ and $g_{3,{\bf{k}},\omega }^a$) flowing in $i$-direction with polarization direction $\alpha$  is proportional to $\nabla _i \dot{\bf{S}}^\alpha  ({\bf{x}},t)$:\cite{Takeuchi}
\begin{eqnarray}
j_{s,i}^{(n)\alpha } ({\bf{x}},t) \propto \nabla _i \dot{{\bf S}}^\alpha  ({\bf{x}},t). 
\end{eqnarray} 
Hence, we can interpret the generation of charge pumping as a result of the coupling of spin current in the normal state $j_{s,l}^{(n)\alpha } ({\bf{x}},t)$ with the triplet vector chirality:
\begin{eqnarray}
j_{c,i}  ({\bf{x}},t) \propto j_{s,l}^{(n)\alpha } ({\bf{x}},t)\sum\limits_{{\bf{k}},\omega } {f'_\omega  } k_i {\mathop{\rm Re}\nolimits} \left[ {\left( {\frac{\partial }{{\partial k_l }}{\bf{f}}_{j,{\bf{k}},\omega }^a  \times {\bf{f}}_{j,{\bf{k}},\omega }^r } \right)^\alpha  } \right] .
\end{eqnarray} 
We see that the triplet vector chirality converts the pumped spin current into charge current.
%Note that here the pumped charge current is generated by the spin structure of triplet pairing. 
Therefore, we do not rely on spin-orbit coupling to generate charge current (convert spin current into charge current) in sharp contrast to the previous works\cite{Saitoh,Valenzuela}.

When the direction of triplet vectors are determined only by the wavevector ${\bf{k}}$, one can decompose the triplet vectors as \cite{comment2} 
\begin{eqnarray}
{\bf{f}}_{j,{\bf{k}},\omega }^a  = f_{j,{\bf{k}},\omega }^a {\bf{n}}_{j,{\bf{k}}} 
\end{eqnarray}
with 3D unit vectors ${\bf{n}}_{j,{\bf{k}}} (j=1,2)$. The pumped charge current can then be written as 
\begin{eqnarray}
j_{c,i} ({\bf{x}},t) = \frac{{4\hbar ^2 eJ}}{{Vm}}\nabla _l \dot{\bf{S}}^\alpha  ({\bf{x}},t) \nonumber \\
\times  \sum\limits_{{\bf{k}},\omega }{f'_\omega  } k_i \left| {f_{j,{\bf{k}},\omega }^a } \right|^2 \left( {\frac{\partial }{{\partial k_l }}{\bf{n}}_{j,{\bf{k}}}  \times {\bf{n}}_{j,{\bf{k}}} } \right)^\alpha. \label{jc2}
\end{eqnarray}
We find that, to obtain nonzero current, $\frac{\partial }{{\partial k_l }}{\bf{n}}_{j,{\bf{k}}} \ne {\bf{0}}$ is necessary. This condition means that the direction of the triplet vector depends on the wavevector ${\bf{k}}$. Namely, rotating the triplet vector induces the change of ${\bf{k}}$, which indicates that dynamics of the triplet vector leads to that of the Cooper pair. Therefore, this condition confirms the intuitive picture we have presented above.

To estimate the current, we consider transparent interface between ferromagnet and triplet superconductor such that the proxmity effect is sufficiently strong and the Green's function in the ferromagnet has the same form as that in the bulk superconductor:
\begin{eqnarray}
f_{j,{\bf{k}},\omega }^a  = \frac{{d_{j,{\bf{k}}} }}{{(\omega  - i\gamma )^2  - \xi ^2  - \Delta _{\bf{k}}^2 }},
\end{eqnarray}
with $\Delta _{\bf{k}}^2  = (d_{1,{\bf{k}}} )^2  + (d_{2,{\bf{k}}} )^2$ where $\gamma$ is the impurity scattering rate. 
Then, the charge current can be estimated as 
\begin{eqnarray}
j_{c,i} ({\bf{x}},t)  \cong  - \frac{{\hbar ^2 e \nu J}}{{Vm}}\nabla _l \dot{\bf{S}}^\alpha  ({\bf{x}},t)  \nonumber \\
\times \left\langle {\frac{{k_i } {(d_{j,{\bf{k}}} )^2 }}{{(\gamma ^2  + \Delta _{\bf{k}}^2 )^{3/2} }}\left( {\frac{\partial }{{\partial k_l }}{\bf{n}}_{j,{\bf{k}}}  \times {\bf{n}}_{j,{\bf{k}}} } \right)^\alpha  } \right\rangle _{FS} 
\end{eqnarray}
at zero temperature where $\nu$ is the density of states at the Fermi level and  $\left\langle ...  \right\rangle _{FS}$ means the average over the Fermi surface.  As an example, let us consider helical $p$-wave superconductivity as $d_{1,{\bf{k}}}  = \Delta _0^{}, {\bf{n}}_{1,{\bf{k}}}  = 1/\left| {\bf{k}} \right|(k_y , - k_x ,0), d_{2,{\bf{k}}}  = 0,$ and $ {\bf{n}}_{2,{\bf{k}}}  = {\bf{0}}$ which would be realized in non-centrosymmetric superconductor. \cite{Frigeri} 
%Note that in general, singlet and triplet pairings are mixed in non-centrosymmetric superconductor, but singlet pairing do not contribute to the current since the contribution from the singlet pairing vanishes by taking trace over spin space in Eq.(\ref{jc}). 
Then, assuming cylindrical Fermi surface, we have the pumped current flowing in $x$-direction of the form 
\begin{eqnarray}
j_{c,x} ({\bf{x}},t)  = \frac{{\hbar ^2 e \nu J}}{{2mV\Delta _0^{} }}\nabla _y \dot{\bf{S}}^z ({\bf{x}},t) \label{jcex}
\end{eqnarray}
in the limit of $\gamma  \to 0 $. We see that the pumped charge current flows perpendicularly to the pumped spin current (note $j_{s,y}^z ({\bf{x}},t) \propto \nabla _y {\dot {\bf{S}}}^z ({\bf{x}},t)$) as illustrated in Fig. \ref{fig1}. 
As an example of the ferromagnet, we consider conical ferromagnet characterized by ${\bf{S}}({\bf{x}},t) = S(a,\cos ({\bf{q}} \cdot {\bf{x}} - \Omega t),\sin ({\bf{q}} \cdot {\bf{x}} - \Omega t))$ where ${\bf{q}}$ is a 3D vector which determines the spin structure, and $a$ and $\Omega$ are real constants. The charge current  flowing in $x$-direction then becomes 
\begin{eqnarray}
j_{c,x} ({\bf{x}},t) = \frac{{\hbar ^2 e \nu J}}{{2mV\Delta _0^{} }}S\Omega q_y \sin ({\bf{q}} \cdot {\bf{x}} - \Omega t).
\end{eqnarray}

%%%%%%%%%%%%%%%%%%%%%%%%%%%%%%%%%%%%%%%%%%%%%%%%%%%%%%%%%%%%%%%%%%%%%%%%%%%%%%%%%%%%%

Now, let us comment on the relevance of our results to SU(2) gauge field.
It has been shown that  the charge and spin currents in inhomogeneous magnetic structure with dynamical spin are associated with SU(2) gauge field. \cite{Volovik,Tatara}
In analogy with the SU(2) gauge field in inhomogeneous magnetic structure with dynamical spin, we introduce the SU(2) gauge field by triplet spin structure.
Then, the component of the SU(2) gauge field perpendicular to the adiabatic component ${\bf{n}}_{\bf{k}}$ reads\cite{Volovik,Tatara}
\begin{eqnarray}
{\bf{A}}_{j,{\bf{k}},\mu }^ \bot   = \frac{1}{2}{\bf{n}}_{j,{\bf{k}}}  \times \frac{\partial }{{\partial k_\mu  }}{\bf{n}}_{j,{\bf{k}}} .
\end{eqnarray}
Hence, we can represent the pumped charge current in terms of the SU(2) gauge field projected  on U(1) as 
\begin{eqnarray}
j_{c,i} ({\bf{x}},t) =  - \frac{{8 \hbar ^2 eJ}}{{Vm}}\nabla _l \dot{\bf {S}}^\alpha  ({\bf{x}},t) \nonumber \\
\times \sum\limits_{{\bf{k}},\omega } {f'_\omega  } k_i \left| {f_{j,{\bf{k}},\omega }^a } \right|^2 \left( {{\bf{A}}_{j,{\bf{k}},l}^ \bot  } \right)^\alpha. 
\end{eqnarray}
We see that the charge current is driven by the gauge field. However, since this current is also driven by dynamical spins, it has a dissipative nature (it is not a supercurrent).
% and this equation can be regarded as "the London equation". 
In the strong coupling limit, only the adiabatic component of the SU(2) gauge field, namely that projected on ${\bf{n}}_{\bf{k}}$, couples to the charge. \cite{Volovik}
In contrast, here the component perpendicular to ${\bf{n}}_{\bf{k}}$ contributes to the charge current. \cite{Tatara,Taguchi2}

In this paper, we have considered charge pumping in ferromagnet/triplet superconductor junctions. However, our results are also applicable to ferromagnetic superconductors where ferromagnetism and triplet superconductivity coexist in the bulk. 
If a ferromagnetic superconductor has some inhomogeneity such as  domain wall structure, charge pumping would occur by applying ac magnetic field.

%Recent experimental progress has made it possible to fabricate inhomogeneous ferromagnet/$s$-wave superconductor junctions.\cite{Keizer,Khaire,Robinson} We hope that inhomogeneous ferromagnet/triplet superconductor junctions will be fabricated. 

In summary,
 we have investigated charge pumping in ferromagnet/triplet superconductor junctions where the magnetization of the ferromagnet is inhomogeneous and dynamical. It has been shown that charge current is pumped due to the coupling of the localized spins with triplet vector chirality. This can be also interpreted as a result of the coupling between the pumed spin current and the  triplet vector chirality. 
Physical mechanism of the charge pumping has been also presented. 
%This mechanism  does not rely on spin-orbit coupling in contrast to the previous one originating from magnetization dynamics\cite{Saitoh,Valenzuela}. 

This work was supported by Grant-in-Aid for Young Scientists (B) (No. 23740236) and the "Topological Quantum Phenomena" (No. 23103505) Grant-in Aid for Scientific Research on Innovative Areas from the Ministry of Education, Culture, Sports, Science and Technology (MEXT) of Japan.

\end{document}